\documentclass[conference]{IEEEtran}
\IEEEoverridecommandlockouts
\usepackage{cite}
\usepackage{amsmath,amssymb,amsfonts}
\usepackage{algorithmic}
\usepackage{graphicx}
\usepackage{booktabs}
\usepackage[greek,american]{babel}
\usepackage{tabularx}
\usepackage{svg}

\usepackage{textcomp}
\usepackage{hyperref}
\usepackage{xcolor}
\def\BibTeX{{\rm B\kern-.05em{\sc i\kern-.025em b}\kern-.08em
    T\kern-.1667em\lower.7ex\hbox{E}\kern-.125emX}}

\usepackage[font=footnotesize, labelsep=period]{caption}
\begin{document}

\title{Reflections on Diversity: A Real-time Virtual Mirror for Inclusive 3D Face Transformations}


\author{\IEEEauthorblockN{Paraskevi Valergaki}
\IEEEauthorblockA{\textit{Computer Science Department,} \\
\textit{University of Crete}\\
\IEEEauthorblockA{\textit{Institute of Computer Science,} \\
\textit{Foundation for Research \& Technology - Hellas (FORTH)}\\
Heraklion, Greece \\
vivibalergaki@gmail.com}}
\and
\IEEEauthorblockN{Antonis Argyros}
\IEEEauthorblockA{\textit{Computer Science Department,} \\
\textit{University of Crete}\\
\IEEEauthorblockA{\textit{Institute of Computer Science,} \\
\textit{Foundation for Research \& Technology - Hellas (FORTH)}\\
Heraklion, Greece \\
argyros@ics.forth.gr}}
\and
\IEEEauthorblockN{Giorgos Giannakakis\textsuperscript{*}}
\IEEEauthorblockA{\textit{Institute of Computer Science,} \\
\textit{Foundation for Research \& Technology - Hellas (FORTH)}\\
Heraklion, Greece \\
\textit{Department of Electronic Engineering,} \\
\textit{Hellenic Mediterranean University}\\
Chania, Greece \\
ggian@ics.forth.gr}
\and
\IEEEauthorblockN{Anastasios Roussos\textsuperscript{*}}
\IEEEauthorblockA{\textit{Institute of Computer Science,} \\
\textit{Foundation for Research \& Technology - Hellas (FORTH)}\\
Heraklion, Greece \\
troussos@ics.forth.gr}

\thanks{\textsuperscript{*} Joint last authorship.}
}

\maketitle

\begin{abstract}
Real-time 3D face manipulation has significant applications in virtual reality, social media and human-computer interaction. This paper introduces a novel system, which we call Mirror of Diversity (MOD), that combines Generative Adversarial Networks (GANs) for texture manipulation and 3D Morphable Models (3DMMs) for facial geometry to achieve realistic face transformations that reflect various demographic characteristics, emphasizing the beauty of diversity and the universality of human features. As participants sit in front of a computer monitor with a camera positioned above, their facial characteristics are captured in real time. Our system provides a dynamic, responsive ``mirror'' effect, allowing the digital 3D model to follow the participant’s motions, offering an immersive virtual reflection.
Participants can further alter their digital face reconstruction with transformations reflecting different demographic characteristics—such as gender and ethnicity (e.g., a person from Africa, Asia, Europe). 
Another feature of our system, which we call ``Collective Face'', generates an averaged face representation from multiple participants’ facial data.
A comprehensive evaluation protocol is implemented to assess the realism
and demographic accuracy of the transformations. Qualitative feedback
is gathered through participant questionnaires, which include comparisons
of MOD transformations with similar filters on platforms like Snapchat and TikTok, focusing on realism, feature preservation, and faithfulness to demographic representation. Additionally, quantitative analysis is conducted using a pretrained Convolutional Neural Network that predicts gender and ethnicity, to validate the accuracy of demographic transformations. Project webpage: \url{https://vivianval.github.io/ReflectionsOnDiversity}

\end{abstract}
\section{Introduction}
The intersection of artificial intelligence, computer graphics, and human-computer interaction has witnessed rapid advancements over recent years, driven largely by the development of deep learning architectures like Generative Adversarial Networks (GANs) and 3D Morphable Models (3DMMs). Within this context, the ability to manipulate 3D face representations in real time is particularly compelling. Traditional 3D face modeling techniques, while accurate, often lack the flexibility to produce the full range of realistic textures that users demand. Similarly, GAN-based face generators, though capable of producing highly realistic 2D images, are not inherently designed for 3D face manipulation and struggle to maintain identity consistency. What is more, existing face manipulation applications~\cite{bbc_asian_filter, bbc_bob_marley, bbc_coachella_filter, menderthal}, rarely address transformations based on ethnicity. Additionally, to the best of our knowledge, no face manipulation application provides a tool for users to explore realistic, ethnicity-based transformations in a meaningful way. Our approach aims to address the aforementioned limitations along with the core motivation to develop technology that fosters inclusiveness and diversity. 

In this paper, we present the Mirror of Diversity (MOD) computer vision software application, as part of the \textit{STEAMDIVE} project, which allows users to experience realistic transformations of their faces as if they were of a different gender or ethnicity. This educational tool is in the form of a virtual interactive mirror that raises awareness around identity, multiculturalism, promoting inclusion and diversity in an educational environment.

\section{Related Work}
The concept of virtual mirror has been utilized in various behavioural \cite{grewe2023open},  psychological \cite{inoue2021virtual} and healthcare applications \cite{andreu2015mirror,andreu2016wize}. 

\begin{figure}[t]
\centering
\includegraphics[width=1.0\linewidth]{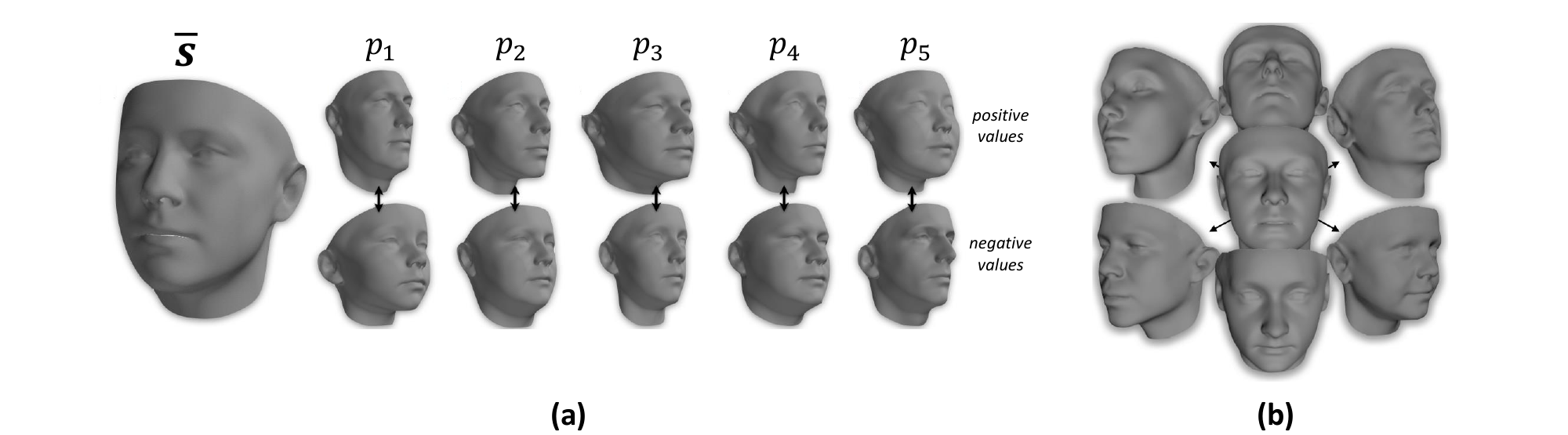}
\caption[Linear shape identity model of the global LSFM model]{\label{figure:relatedwork1}
Visualization of the linear shape identity model of the global LSFM model: \textbf{(a)} Mean shape ($\bar{\textbf{s}}$) and first 5 basis shapes (out of a total of $n_{id}$=158 basis shapes), each visualized as additions
and subtractions away from the mean shape. In more detail, the top (bottom) row corresponds to setting 
the weight $p_i$ of the $i$-th basis shape to a positive (negative) value, corresponding to 3 standard deviations of its statistical distribution. 
\textbf{(b)} Generation of synthetic shapes through random sampling of the shape coefficients $\textbf{p}$, assuming a Gaussian distribution. Figure adapted from \cite{booth_large_2018}.}
\end{figure}

\noindent\textbf{Large Scale Facial Models:}  
Booth et al.~\cite{booth_large_2018} introduced the \textit{Large Scale Facial Model (LSFM)}, a 3D Morphable Model (3DMM) trained on 12,000 high-resolution 3D scans (see Fig.~\ref{figure:relatedwork1}). Compared to previous models, such as the Basel Face Model~\cite{5279762}, LSFM exhibits greater facial variation and less tight constraints.

As 3DMMs, LSFMs consist of three components: shape identity, blendshapes, and color model. The \textit{shape model} assumes a fixed topology \( T \), modeling vertex coordinates as \( s_{\text{id}}(p) = \bar{s} + Up \), where \( \bar{s} \) is the mean shape, and \( U \) contains identity basis shapes, learned via PCA. The \textit{blendshapes model} extends identity to capture expressions as \( s(p, q) = \bar{s} + Up + Vq \), where \( V \) encodes expression variations. Similarly, the \textit{color model} follows \( c(\lambda) = \bar{c} + W\lambda \), where \( W \) contains color basis vectors.

To construct LSFM, 3D data was captured using a 3dMD™ stereo photometric device. Face meshes were aligned to establish dense correspondence, ensuring that each vertex encodes the same semantic point across samples. While \textit{triangle meshes} are the standard representation, alternatives include \textit{cylindrical}~\cite{6795900}, \textit{orthographic}~\cite{10.1007/978-3-540-24671-8_8}, and \textit{per-vertex surface normals}~\cite{10.1007/978-3-642-33712-3_15}.

PCA was applied, retaining 158 principal components, explaining 99.7\% of variance. LSFM captures identity variations across multiple demographics, enhanced by the \textit{FaceWarehouse dataset}~\cite{6654137}, which provides expression variations from 150 individuals. As high-dimensional facial shape and texture variations naturally align with demographic features, custom models were derived from MeIn3D and trained on specific demographic subsets using available data (48\% male, 52\% female; 82\% White, 9\% Asian, 5\% mixed heritage, 3\% Black, 1\% other). 



\vspace*{0.2cm}\noindent\textbf{3DMM Fitting Approaches:}
3DMM fitting methods primarily follow two approaches: \textit{analysis-by-synthesis} and \textit{deep learning}. 
In analysis-by-synthesis, the model parameters are adapted in an iterative manner until the synthesized shape matches
the input data. Blanz and Vetter~\cite{10.1145/311535.311556} pioneered an analysis-by-synthesis approach using iterative optimization techniques (e.g., gradient descent, Gauss-Newton). The goal is to estimate an optimal parameter vector \( P^* \) minimizing \( E(P) \), which integrates image error, shape, texture, and rendering priors.  Follow-up works framed the problem as a more general non-linear optimization, see e.g.~ \cite{li2013realtime, cao2014displaced, garrido2016reconstruction, Thies_real_time, face2face, thies2016facevr}.

Deep learning (DL) methods can be supervised or unsupervised. Supervised approaches train networks (e.g., regressors or encoders) to predict 3DMM coefficients from images~\cite{Olszewski, laine2017productionlevelfacialperformancecapture, kim2018inverse, Koujan_2020_CVPR, Doukas_2021}. Unsupervised methods, such as \cite{DBLP:journals/corr/abs-1806-06098, tewari17MoFA, deng2019accurate, on-learning-3d-face-morphable-model-from-in-the-wild-images, tewari2020complete}, are more flexible since they eliminate the need for 3D ground truth annotations.

\vspace*{0.2cm}\noindent\textbf{GANs in Face Generation}:
Early GANs, such as Goodfellow et al.~\cite{goodfellow2014generativeadversarialnetworks}, generated low-resolution grayscale images. By 2018, advances like Deep Convolutional GANs (i.e. DCGANs) ~\cite{radford2016unsupervisedrepresentationlearningdeep},Wasserstein GANs ~\cite{arjovsky2017wassersteingan}, and StyleGAN ~\cite{karras2019stylebasedgeneratorarchitecturegenerative} enabled the synthesis of high-quality photorealistic faces. Progressive GANs (ProGANs) ~\cite{karras2018progressivegrowinggansimproved} introduced hierarchical training, stabilizing GAN convergence by growing both generator and discriminator simultaneously from low to high resolution. Their CelebA-HQ dataset significantly improved photorealism and serves as a reference in this study. 

\textit{FacialGAN} ~\cite{durall2021facialganstyletransferattribute} performs style transfer while preserving identity, utilizing semantic segmentation masks to control attributes like eyes, lips, and skin. It supports interactive editing~\cite{lee2020maskgandiverseinteractivefacial}, allowing real-time adjustments. The framework consists of Generative, Style, Segmentation, and Discriminative Networks. 


\begin{figure*}
    \centering
    \includegraphics[width=1\linewidth]{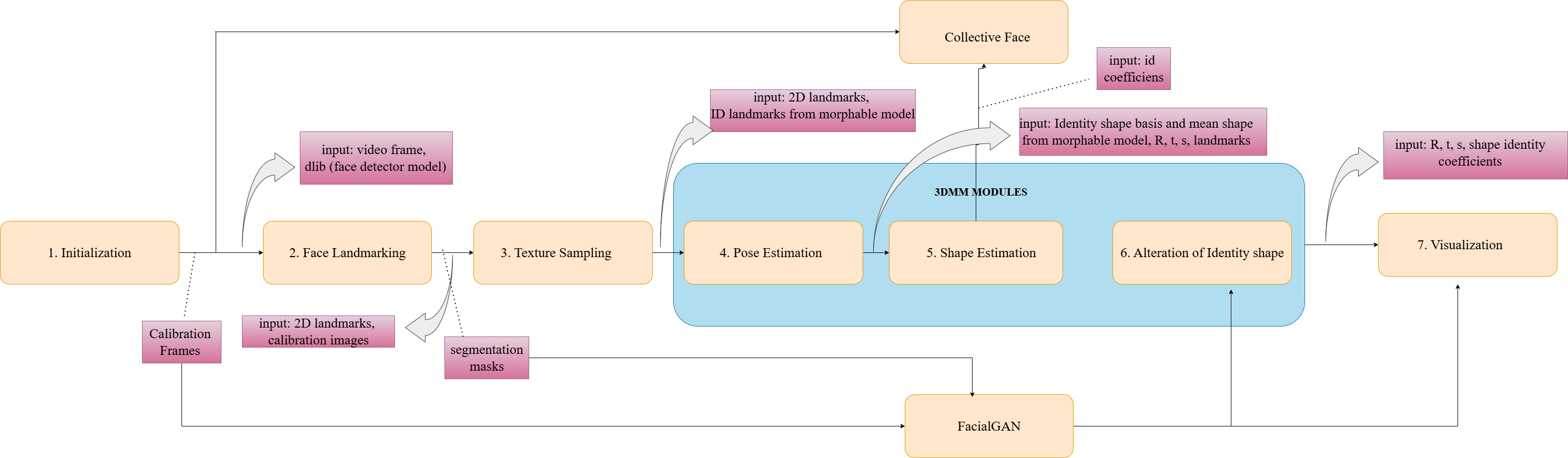}
    \caption{Pipeline of the proposed system.}
    \label{fig:flowchart}
\end{figure*}

\section{Methods}

Figure~\ref{fig:flowchart} presents an overview of the proposed pipeline. It consists of seven modules. Once the calibration frames are captured, face landmarking extracts the landmarks and segmentation masks. The latter are input to FacialGAN which will produce the texture corresponding to the user's options in terms of transformations. 3DMM modules are responsible for fitting, pose and shape estimation or alteration of identity shape and therefore they are calculated at each frame. At this stage, identity coefficients along with the frontal calibration frame are sent to Collective Face. In the following sections, we provide more details about the main modules of our pipeline.

\subsection{3DMM Fitting}\label{AA}
Our application employs 3DMM fitting for face reconstruction by aligning pre-constructed models with 2D facial landmarks and optionally using depth maps and texture data to generate a 3D face that closely matches the input.

3DMMs are modeled as a linear combination of identity and expression variations~\cite{10.1007/s11263-018-1134-y,10.1109/TPAMI.2018.2832138}. We use LSFM models (global or bespoke), which incorporate demographic metadata~\cite{booth_large_2018}. Fitting involves estimating face pose and shape using an orthographic camera model, aligning extracted landmarks with their 3DMM correspondences, and solving a linear system via Singular Value Decomposition (SVD). More specifically, the 3D shape is projected into 2D by applying 3D rotation and translation matrices, which account for the camera's position, and then the vertices are mapped to 2D based on factors like focal length and perspective. Identity and expression coefficients are estimated by minimizing a linear least squares problem, similar to~\cite{10.1007/s11263-018-1134-y}. To mitigate jitter, a temporal smoothing step averages poses across three consecutive frames.

\vspace*{0.1cm}\noindent\textit{Pose Estimation:}  
Pose estimation determines head position via a rotation matrix \( R \), translation vector \( t \), and scale \( s \). Following~\cite{BASANILSMITH}, the method constructs a linear system where matrix \( A \) encodes the 3D model points, and vector \( d \) contains 2D landmarks:
\begin{align}
A \mathbf{k} &= d, \quad \text{where }  \\
A_{2i-1} &= 
\begin{bmatrix} u_i & v_i & w_i & 1 & 0 & 0 & 0 & 0 \end{bmatrix}, \\
A_{2i} &= 
\begin{bmatrix} 0 & 0 & 0 & 0 & u_i & v_i & w_i & 1 \end{bmatrix}.
\end{align}

Solving for \( \mathbf{k} \) yields the translation components \( t_x = k_3 / s, \, t_y = k_7 / s \) and scale \( s = (\|\mathbf{R}_1\|_2 + \|\mathbf{R}_2\|_2)/2 \). The 3D rotation matrix is refined using Singular Value Decomposition (SVD). 
For temporal stability, smoothing is applied to pose estimation. The rotation matrix is updated using a weighted combination of the current and the previous frame, ensuring smoother transitions. 

\vspace*{0.1cm}\noindent\textit{Shape Estimation:}
The goal is to estimate face shape coefficients by fitting a 3D morphable model to detected 2D landmarks. Given pose estimation parameters (\textit{R}, \textit{s}, \textit{t}), mean shape $\bar{f}$, and principal components $P$, the process involves (1) constructing cumulative matrices $A = \text{proj} \cdot P$, $h = \text{landmarks} - s \cdot t - \text{proj} \cdot \bar{f}$, with $\text{proj} = s \cdot (I \otimes R_{[1:2]})$; (2) applying hyper-box constraints with $C = [I \ -I]$ and $d = \pm k \cdot \lambda$ so that we come up with plausible faces, (3) minimizing reprojection error $\|A \cdot x - h\|^2$ subject to $C \cdot x \leq d$, and (4) extracting identity coefficients $\alpha$ and reconstructing shape $\hat{f} = P \cdot \alpha + \bar{f}$.
It should be mentioned that, for the sake of robustness, our algorithm considers only identity shape parameters during the calibration phase.

\begin{figure}
    \centering
    \includegraphics[width=0.7\linewidth]{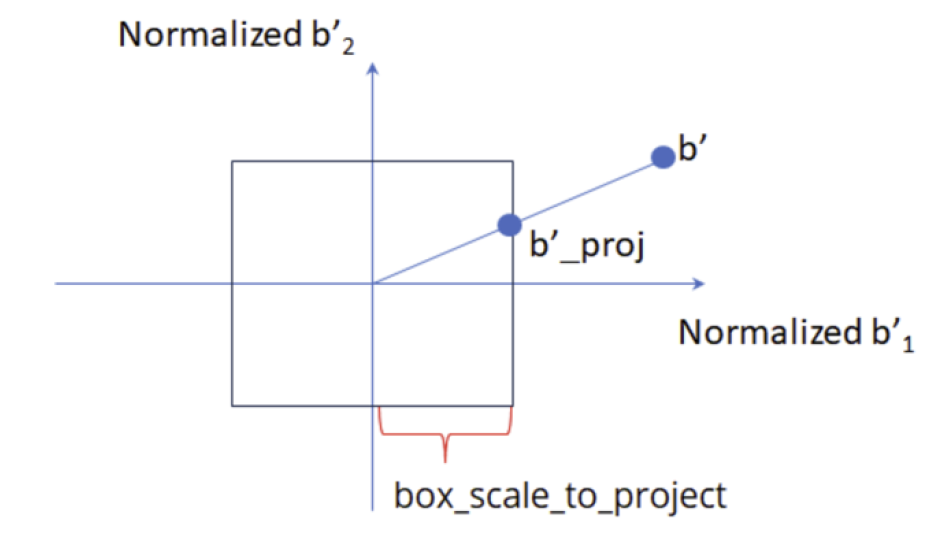}
    \caption{PCA projection of identity shape into the allowable range defined by box\_scale\_to\_project}
    \label{fig:pcaproj}
\end{figure}

\vspace*{0.1cm}\noindent\textit{Alteration of Identity shape:}
The goal is to transform a participant's 3D facial geometry to resemble a target demographic. Given identity shape $s$, the bespoke model's mean shape $\mu'$, principal components $U'$, and standard deviations $\sigma$, the pipeline involves: (1) centering the shape as $s_{\text{centered}} = s - \mu'$; (2) projecting onto PCA space as $b' = s_{\text{centered}}^T U'$; (3) applying constraints via $b'_{\text{proj}} = b' \cdot \frac{\text{box\_scale\_to\_project}}{\max(|b'|/\sigma)}$ if $\max(|b'|/\sigma) > \text{box\_scale\_to\_project}$; (4) reconstructing the altered shape as $s_{\text{PCAproj}} = U' b'_{\text{proj}} + \mu'$ (see Figure \ref{fig:pcaproj}).

\subsection{Morphing}
Morphing interpolates between two 3D face states over time, transitioning from the original to a PCA-constrained version. The interpolation factor \( p_{\text{morph}}(t) \) follows a sinusoidal function \[
p_{\text{morph}}(t) = \frac{1 + \sin \left( 2 \pi \frac{t}{T_{\text{morphing}}} \right)}{2},
\] where \( T_{\text{morphing}} \) is the total morphing period. This ensures smooth oscillation between 0 and 1, controlling the blend between the original and projected shapes. Morphing updates occur periodically based on a frame counter mechanism, ensuring seamless transitions.

\subsection{FacialGAN}

To modify facial textures, we build upon FacialGAN ~\cite{durall2021facialganstyletransferattribute}. We adopt the MULTI-PIE markup scheme~\cite{10.1016/j.imavis.2009.08.002} with 68 facial landmarks (see Fig. 4(a)).

The alignment process begins by computing the eye centers, where the left eye center is \( \mathbf{C}_{\text{left}} = \frac{1}{6} \sum_{i=37}^{42} (x_i, y_i) \) and the right eye center is \( \mathbf{C}_{\text{right}} = \frac{1}{6} \sum_{i=43}^{48} (x_i, y_i) \). The rotation angle is then calculated as \( \theta = \arctan \left( \frac{y_{\text{right}} - y_{\text{left}}}{x_{\text{right}} - x_{\text{left}}} \right) \), while the scale factor is determined as \( s = \frac{d_{\text{desired}}}{\sqrt{(x_{\text{right}} - x_{\text{left}})^2 + (y_{\text{right}} - y_{\text{left}})^2}} \), ensuring proper spatial consistency.

Using these parameters, we calculate a rigid transformation $M$ to bring all faces in a common reference space:
\[
\mathbf{M} =
\begin{bmatrix}
s\cos\theta & -s\sin\theta & t_x \\
s\sin\theta & s\cos\theta  & t_y
\end{bmatrix}
\]
and apply it to the original image \( I \), yielding the aligned face \( I_{\text{aligned}} \). Similarly, each landmark \( \mathbf{p}_i = (x_i, y_i) \) is transformed using \( \mathbf{p}_i' = \mathbf{M} \cdot \mathbf{p}_i \).

To generate segmentation masks for distinct facial regions (e.g., nose, mouth, eyes), we use the transformed facial landmarks as well as a UNet-based segmentation ~\cite{papantoniou2022neuralemotiondirectorspeechpreserving} (see Fig. 4(b)). The final aligned and segmented face and masks are sent to FacialGAN, producing high-quality \( 256 \times 256 \) transformed images. Reference images are selected from CelebA-HQ~\cite{karras2018progressive} for each bespoke transformation. According to the modification type determined by the user, an interpolation function for smooth texture blending that combines manipulated and non-manipulated textures is applied.


\begin{figure}
    \centering
    \includegraphics[width=1\linewidth]{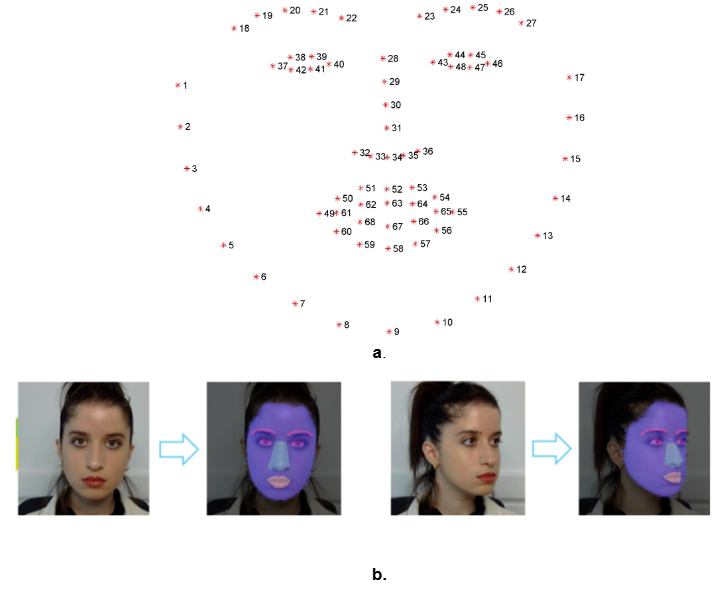}
    \caption{(a) Adopted mark-up scheme of 68 facial landmarks (Figure from ~\cite{10.1007/s11263-018-1134-y}), (b) The process of mask generation. The first image shows the input face, the second illustrates the segmented face with different regions (mouth, nose, eyes, and eyebrows) covered by distinct masks and visualized by different colors.}
    \label{fig:landmarksfgsde}
\end{figure}

\subsection{Collective Face}

The \textit{Collective Face} feature generates an averaged face from multiple participants' data. Facial landmarks are extracted, aligned, and identity coefficients are averaged for continuous updates. The collective face is periodically sent to FacialGAN via Dropbox for texture or expression manipulation. A Graphical User Interface (GUI) enables users to assign their transformation to Collective Face F (female) or Collective Face M (male), with future support planned for non-binary options.

\subsection{Implementation}

\begin{figure}
    \centering
    \includegraphics[width=0.9\linewidth]{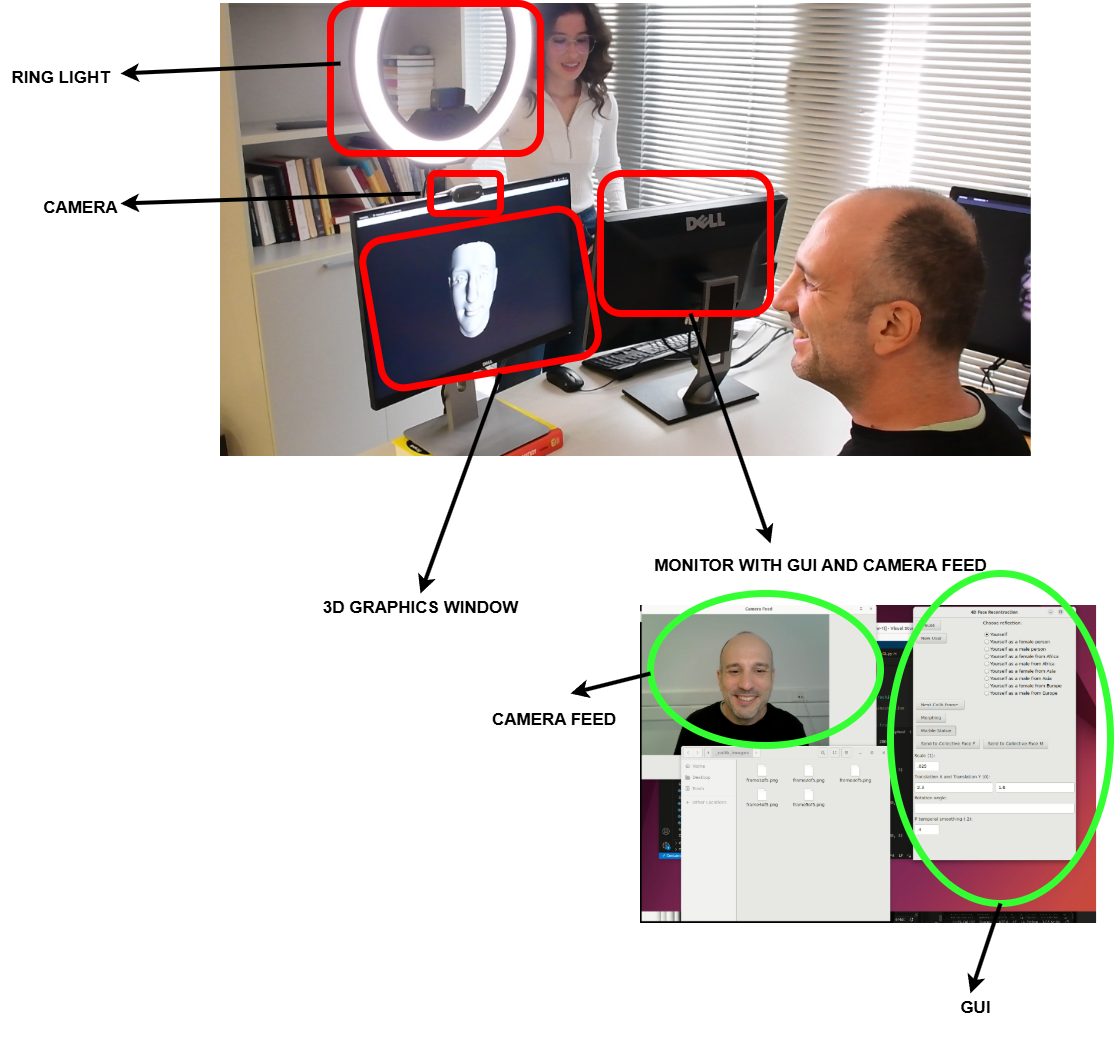}
    \caption{Installation setup with GUI, camera, lighting, and OpenGL rendering.}
    \label{fig:installation}
\end{figure}

The installation includes a computer, two monitors (one for the instructor, one for the participant), a Logitech C270 HD webcam, and a ring light for uniform illumination. The system processes the live feed and mirrors it for the virtual reflection effect. The GUI allows real-time calibration, adjusting scaling and translation parameters for precise alignment (Figure~\ref{fig:installation}).

Our system was tested on an AMD Ryzen 9 7900 (12-core, 24-thread) processor, 61 GiB RAM, and NVIDIA RTX 3060 for GPU-accelerated processing. The system maintains a frame rate of 60 FPS as reported by the OpenGL rendering loop.

\section{Experiments}
The evaluation of our Mirror Of Diversity system's face reconstruction and ethnicity transformations is conducted through a combination of quantitative and qualitative methods, aiming to assess both the technical accuracy and the user-perceived realism and accuracy of the transformations.

\subsection{Qualitative Analysis}

\begin{figure}
    \centering
    \includegraphics[width=0.9\linewidth]{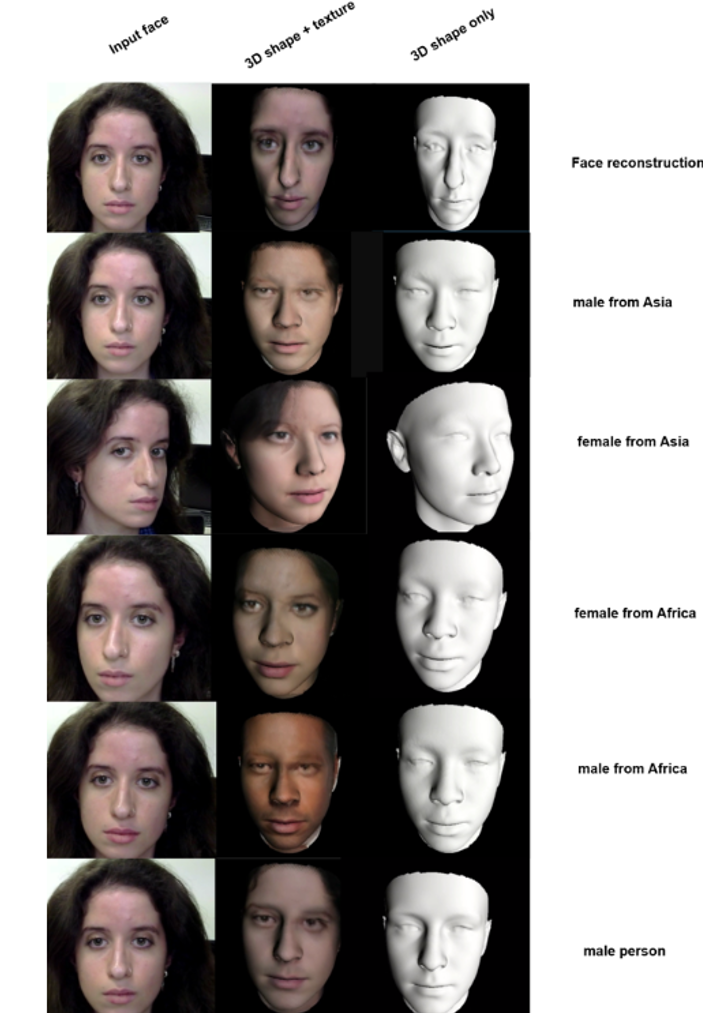}
    \caption{Examples of input and transformed faces using the MOD software. The transformations include face reconstruction, gender-based transformations (male and female), and ethnicity-based transformations (Asian and African). Each transformation is visualized in three formats: the input face, 3D shape with texture, and 3D shape only. A demo video showcasing all the functionalities of our system is available at  \url{https://vivianval.github.io/ReflectionsOnDiversity}.}
    \label{fig:functionality}
\end{figure}

To provide an initial overview of the capabilities of the MOD software, this section presents illustrative examples of its functionalities. Figure \ref{fig:functionality} showcases the capabilities of the MOD software in performing facial transformations across different genders and ethnicities while providing both realistic and structural visualizations. Each row represents a transformation applied to the input face, including gender changes (e.g., male to female) and ethnicity-based modifications (e.g., female from Africa, male from Asia). For each transformation, three outputs are provided: (1) input face, (2) 3D shape with texture, and (3) 3D shape only. The MOD software also supports transformations to European male and female. 

\subsection{Quantitative Evaluation}

\begin{figure}
    \centering
    \includegraphics[width=1\linewidth]{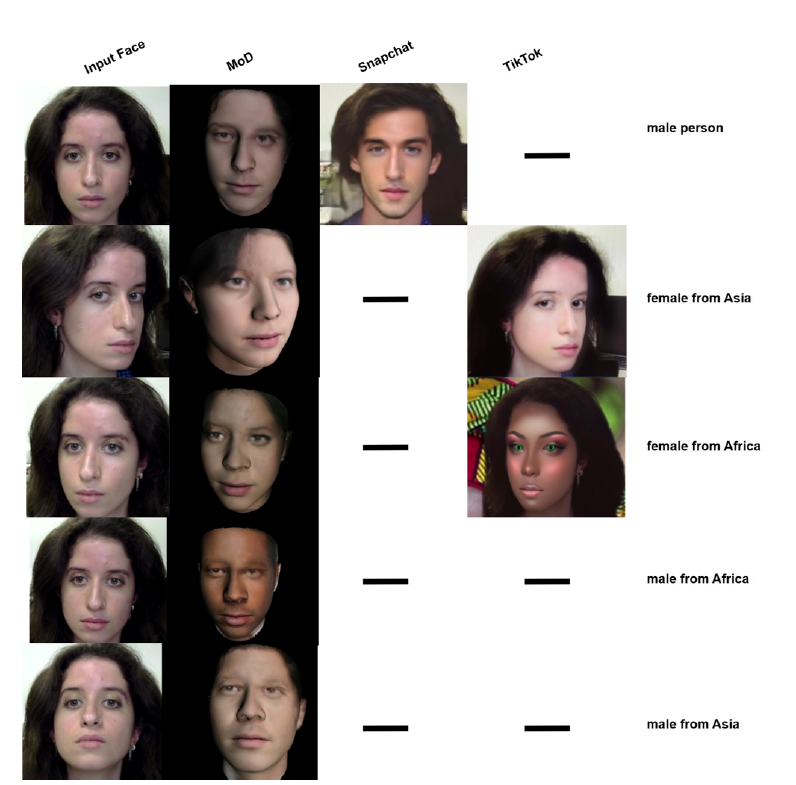}
    \caption{Comparison of how MoD, Snapchat and TikTok handle specific demographic and gender transformations. In each row the first image corresponds to the input original face and then each column is associated with the method used for the transformation. First row presents transformations to a male person achieved with MoD and Snapchat.}
    \label{fig:mobilefilters}
\end{figure}

We evaluate MOD’s transformations using the FairFace model~\cite{kärkkäinen2019fairfacefaceattributedataset}, which employs a pretrained ResNet-34 CNN for gender and ethnicity classification. We compare MOD-generated transformations to face-manipulation filters from Snapchat and TikTok, as well as a version of the Real-Time 4D Face Reconstruction (RT4Dface system)~\cite{koujan2020realtimemonocular4dface}, which was demonstrated during the European Researcher's Night at FORTH, Heraklion, Crete (2019). Figure ~\ref{fig:mobilefilters} showcases manipulations from our system and compared methods.


To assess transformation accuracy, we use FairFace’s classifier to predict gender and ethnicity from OpenGL-rendered MOD images and their counterparts from Snapchat and TikTok. The idea behind this evaluation is that, if a method's face transformation is successful, then an automatic classifier like FairFace should classify the face as belonging to the desired class (e.g. male from Asia). The classifier outputs confidence scores for four ethnicity classes (White, African, Asian, Indian) and gender scores. The dataset includes original and transformed images (gender-swapped, male from Africa, female from Africa, male from Asia, female from Asia). Tables~\ref{tab:gender_classification} - ~\ref{tab:eth_classification} illustrate the acquired metrics. The \textit{macro average} is computed as  $\text{Macro Avg} = \frac{1}{C} \sum_{i=1}^{C} \text{Metric}_i$, treating all classes equally, while the \textit{weighted average} considers class distribution as  $\text{Weighted Avg} = \frac{\sum_{i=1}^{C} (\text{Support}_i \times \text{Metric}_i)}{\sum_{i=1}^{C} \text{Support}_i}$, where $\text{Support}_i$ is the number of samples in class $i$, emphasizing classes with more samples.



\begin{table}[t]
\centering
\scriptsize
\caption{Gender Classification Scores}
\label{tab:gender_classification}
\begin{tabular}{lcc}
\toprule
\textbf{Metric} & \textbf{MOD (Ours)} & \textbf{Mobile Filters} \\
\midrule
F1-Score (Female) & \textbf{0.91} & 0.89 \\
F1-Score (Male) & \textbf{0.92} & 0.67 \\
Accuracy & \textbf{0.92} & 0.83 \\
Macro Avg (F1) & \textbf{0.92} & 0.78 \\
Weighted Avg (F1) & \textbf{0.92} & 0.85 \\
\bottomrule
\end{tabular}
\end{table}

\begin{table}[t]
\centering
\scriptsize
\caption{Ethnicity Classification Scores}
\label{tab:eth_classification}
\begin{tabular}{lcc}
\toprule
\textbf{Metric} & \textbf{MOD (Ours)} & \textbf{Mobile Filters} \\
\midrule
F1-Score (White) & 0.62 & \textbf{0.80} \\
F1-Score (African) & \textbf{0.67} & 0.00 \\
F1-Score (Asian) & 0.40 & \textbf{0.40} \\
Accuracy & \textbf{0.58} & 0.50 \\
Macro Avg (F1) & \textbf{0.56} & 0.40 \\
Weighted Avg (F1) & \textbf{0.56} & 0.40 \\
\bottomrule
\end{tabular}
\end{table}

MOD generates transformed faces that yield higher gender classification accuracy (92\%) compared to mobile filters (83\%), with significantly better male transformation accuracy (F1-Score: 0.92 vs. 0.67). For ethnicity classification, MOD surpasses mobile filters in detecting African and Asian features, though there is area for improvement in these categories.

\subsection{User Study}
We conducted an anonymous online survey with 20 participants to evaluate realism, demographic accuracy, and feature preservation in video transformations. Each participant reviewed 32 video pairs, comparing MOD’s transformations with those from Snapchat, TikTok, and RT4Dface~\cite{koujan2020realtimemonocular4dface}. Transformations were randomized to ensure unbiased evaluations.

Participants rated each transformation on a \textit{5-point Likert scale} based on:
1. \textbf{Demographic Accuracy}: Resemblance to the intended demographic.
2. \textbf{Realism}: How realistic the transformation appears.
3. \textbf{Feature Preservation}: How much the transformation resembles the original person. Weighted averages were calculated to provide a quantitative comparison of user perceptions across the different methods.

\begin{figure}
    \centering
    \includegraphics[width=1\linewidth]{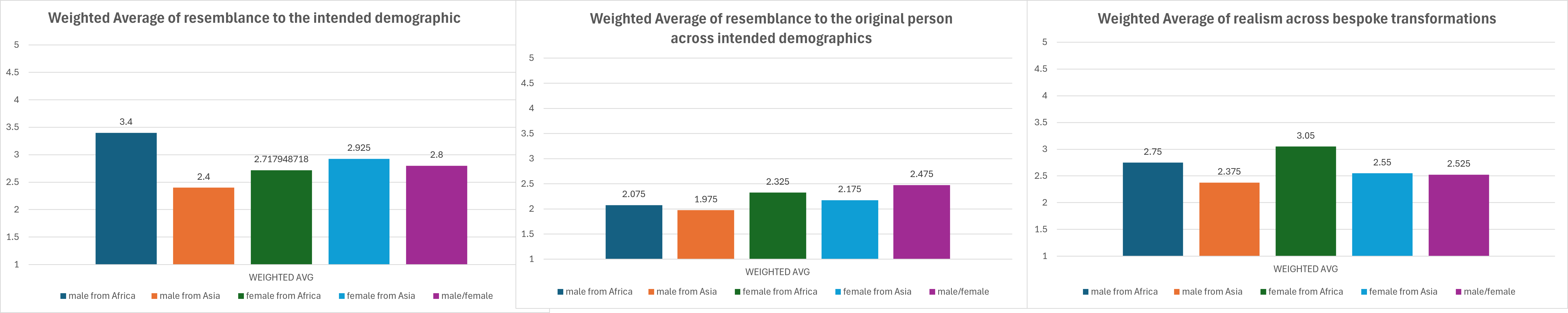}
    \caption{Female and male from Africa have in general higher user ratings in terms of resemblance to intended demographic and realism}
    \label{fig:bespokeanalysis}
\end{figure}

\vspace*{0.1cm}\noindent\textit{Analysis of Bespoke Models:}  
Among demographic transformations, male from Africa scored highest in resemblance (3.4), while male from Asia had the lowest user ratings (2.4). Female from Africa achieved the highest realism (3.05), while male from Asia was the least convincing (2.375). Feature preservation remained moderate, with male/female transformations leading at 2.475 (see Figure~\ref{fig:bespokeanalysis}).

\begin{figure}[htbp]
    \centering
    \includegraphics[width=1\linewidth]{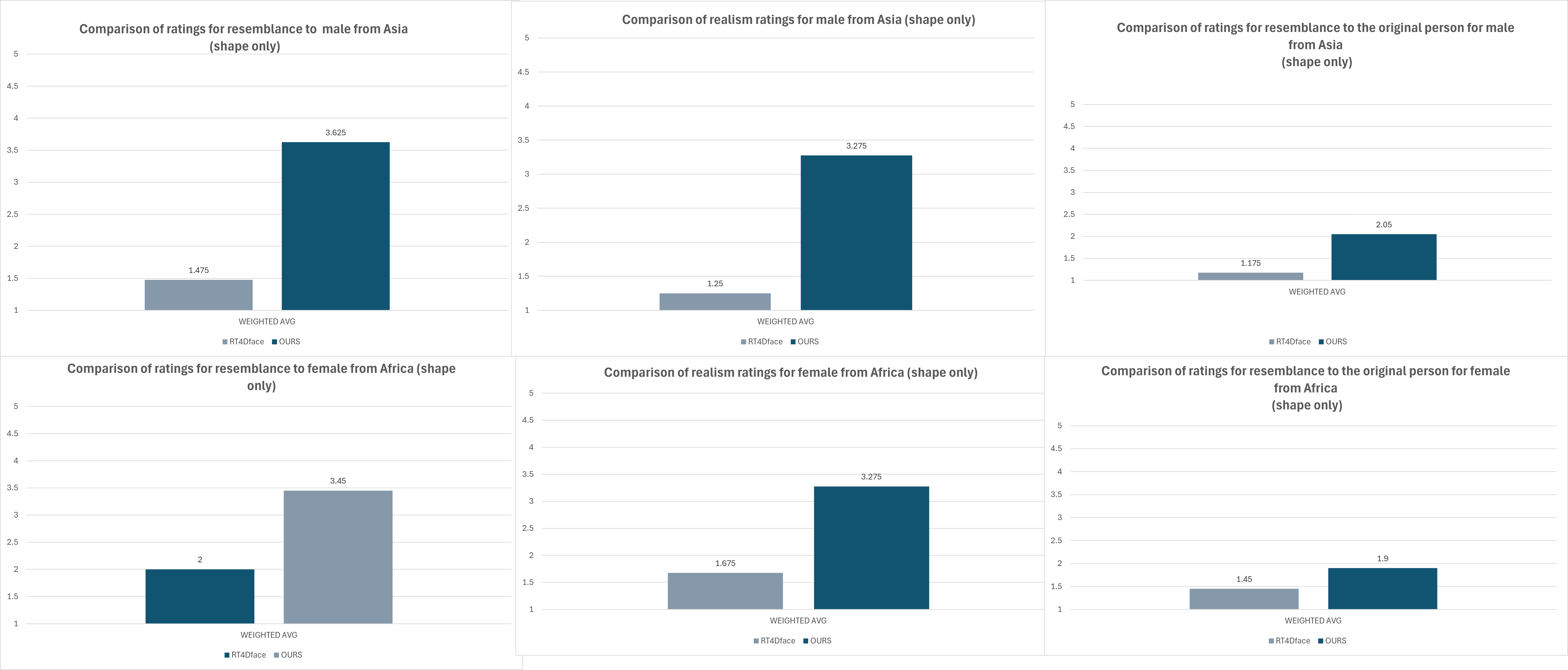}
    \caption{MOD outperforms RT4Dface in realism, demographic resemblance, and identity preservation.}
    \label{fig:surpasses}
\end{figure}

\vspace*{0.1cm}\noindent\textit{Comparative Analysis:}  
MOD significantly outperformed RT4Dface, achieving a weighted average of 3.54 for demographic resemblance (vs. 1.74). The most improved transformation was male from Asia, increasing from 1.5 to 3.55. Realism also improved, scoring 3.27 vs. 1.46 in RT4Dface with a notable shift from lower-rated categories ("Not at all" or "Slightly Accurate") to higher-rated ones ("Moderately Accurate" or above) as depicted in Figure~\ref{fig:surpasses}.


\begin{figure}
    \centering
    \includegraphics[width=1\linewidth]{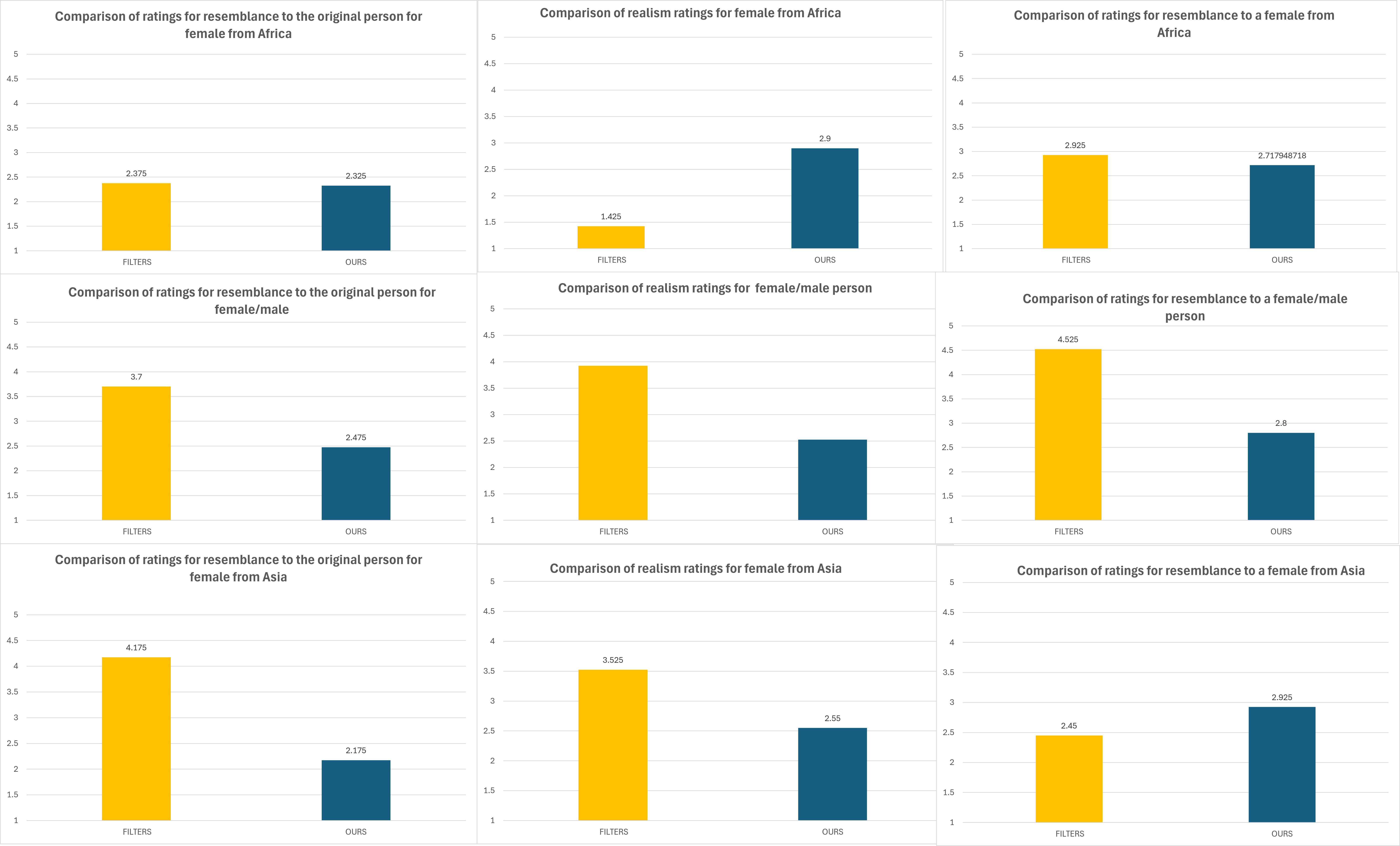}
    \caption{MOD surpasses mobile filters in specific cases as in realism for female from Africa and resemblance to female from Asia.}
    \label{fig:mobflt2}
\end{figure}

\vspace*{0.1cm} Compared to Snapchat and TikTok, MOD’s transformations had lower demographic resemblance (2.8 vs. 3.2) and realism (2.6 vs. 2.9), likely due to real-time expression effects in filters. However, MOD performed better in female from Asia (demographic resemblance) and female from Africa (realism), highlighting its strength in culturally specific transformations (see Figure ~\ref{fig:mobflt2}).

\section{Conclusion and Future Research}

This study demonstrates MOD’s effectiveness in face reconstruction and ethnicity-based transformations, outperforming RT4Dface in realism (3.275 vs. 1.4625) and demographic resemblance (3.537 vs. 1.7375). The system excels in African transformations but struggles with male from Asia, indicating areas for refinement. 

While Snapchat/TikTok generally scored higher, MOD outperformed in specific cases, such as female from Asia (demographic resemblance) and female from Africa (realism). These findings highlight MOD’s strength in fine-grained demographic modeling.
Future improvements should focus on (1)~real-time facial dynamics (blinking, directional gaze) for enhanced realism, (2)~Improved lighting robustness to reduce inconsistencies, (3)~expanding beyond binary gender transformations for greater inclusivity, (4)~refining demographic models to represent more specific cultural identities, (5)~enhancing recall for African and Asian transformations via larger datasets and (6)~ extending demographic transformations to finer-grained cultural representations (e.g., countries instead of continents). These refinements will ensure MOD produces more accurate, diverse, and representative transformations across populations.

\section*{Acknowledgments}

This work was co-funded by the project STEAMDIVE: Diversity in STEAM, Innovative Educational Tools for Promoting Inclusion and Diversity in Schools, supported by the European Commission under the ERASMUS+ Programme, KA220-SCH - Cooperation partnerships in school education (GA: 2022-1-EL01-KA220-SCH-000086968). 

This work was also co-funded by the Hellenic Foundation for Research and Innovation (HFRI) under the ``1st Call for HFRI Research Projects to support Faculty members and Researchers and the procurement of high-cost research equipment'', project I.C.Humans, no 91.


\bibliographystyle{IEEEtran}
\bibliography{bib/paper}





\end{document}